\documentclass[prb,amsmath,amssymb,superscriptaddress,twocolumn]{revtex4}
\usepackage{float}
\usepackage{amsmath}
\usepackage{bibunits}

\usepackage{amssymb}
\usepackage{amsfonts}
\usepackage{euscript}
\usepackage{enumerate}
\usepackage{hhline}
\usepackage{pslatex}
\usepackage{tabularx}
\usepackage[usenames,dvipsnames]{xcolor}

\usepackage{graphicx}

\usepackage{dcolumn}
\usepackage{bm}
\usepackage[sort&compress]{natbib}

\makeatletter
\renewcommand{\@biblabel}[1]{#1. }
\renewcommand{\@dotsep}{500}
\renewcommand{\@pnumwidth}{0em}
\renewcommand{\l@figure}[2]{
\@dottedtocline{1}{1.5em}{2em}{Figure #1}{}\vspace{15pt}}

\newcommand*{\citen}[1]{%
  \begingroup
    \romannumeral-`\x 
    \setcitestyle{numbers}%
    \cite{#1}%
  \endgroup
}
\usepackage[normalem]{ulem}

\begin{document}
\begin{bibunit}

\title{Efficient telecom-to-visible spectral translation through ultra-low power nonlinear nanophotonics}

\author{Xiyuan Lu}\email{xiyuan.lu@nist.gov}
\affiliation{Microsystems and Nanotechnology Division, Physical Measurement Laboratory, National Institute of Standards and Technology, Gaithersburg, MD 20899, USA}
\affiliation{Maryland NanoCenter, University of Maryland,
College Park, MD 20742, USA}
\author{Gregory Moille}
\affiliation{Microsystems and Nanotechnology Division, Physical Measurement Laboratory, National Institute of Standards and Technology, Gaithersburg, MD 20899, USA}
\affiliation{Maryland NanoCenter, University of Maryland,
College Park, MD 20742, USA}
\author{Qing Li}
\affiliation{Microsystems and Nanotechnology Division, Physical Measurement Laboratory, National Institute of Standards and Technology, Gaithersburg, MD 20899, USA}
\affiliation{Maryland NanoCenter, University of Maryland,
College Park, MD 20742, USA}
\affiliation{Electrical and Computer Engineering, Carnegie Mellon University, Pittsburgh, PA 15213, USA}
\author{Daron A. Westly}
\affiliation{Microsystems and Nanotechnology Division, Physical Measurement Laboratory, National Institute of Standards and Technology, Gaithersburg, MD 20899, USA}
\author{Anshuman Singh}
\affiliation{Microsystems and Nanotechnology Division, Physical Measurement Laboratory, National Institute of Standards and Technology, Gaithersburg, MD 20899, USA}
\affiliation{Maryland NanoCenter, University of Maryland,
College Park, MD 20742, USA}
\author{Ashutosh Rao}
\affiliation{Microsystems and Nanotechnology Division, Physical Measurement Laboratory, National Institute of Standards and Technology, Gaithersburg, MD 20899, USA}
\affiliation{Maryland NanoCenter, University of Maryland,
College Park, MD 20742, USA}
\author{Su-Peng Yu}
\affiliation{Time and Frequency Division, Physical Measurement Laboratory, National Institute of Standards and Technology, Boulder, CO 80305, USA}
\affiliation{Department of Physics, University of Colorado, Boulder, CO 80309, USA}
\author{Travis C. Briles}
\affiliation{Time and Frequency Division, Physical Measurement Laboratory, National Institute of Standards and Technology, Boulder, CO 80305, USA}
\affiliation{Department of Physics, University of Colorado, Boulder, CO 80309, USA}
\author{Scott B. Papp}
\affiliation{Time and Frequency Division, Physical Measurement Laboratory, National Institute of Standards and Technology, Boulder, CO 80305, USA}
\affiliation{Department of Physics, University of Colorado, Boulder, CO 80309, USA}
\author{Kartik Srinivasan} \email{kartik.srinivasan@nist.gov}
\affiliation{Microsystems and Nanotechnology Division, Physical Measurement Laboratory, National Institute of Standards and Technology, Gaithersburg, MD 20899, USA}
\affiliation{Department of Physics, University of Maryland,
College Park, MD 20742, USA}
\date{\today}

\begin{abstract}
\noindent The ability to spectrally translate lightwave signals in a compact, low-power platform is at the heart of the promise of nonlinear nanophotonic technologies. For example, a device to link the telecommunications band with visible and short near-infrared wavelengths can enable a connection between high-performance chip-integrated lasers based on scalable nanofabrication technology with atomic systems used for time and frequency metrology.  While second-order nonlinear ($\chi^{(2)}$) systems are the natural approach for bridging such large spectral gaps, here we show that third-order nonlinear ($\chi^{(3)}$) systems, despite their typically much weaker nonlinear response, can realize spectral translation with unprecedented performance. By combining resonant enhancement with nanophotonic mode engineering in a silicon nitride microring resonator, we demonstrate efficient spectral translation of a continuous-wave signal from the telecom band ($\approx$ 1550 nm) to the visible band ($\approx$ 650 nm) through cavity-enhanced four-wave mixing. We achieve such translation over a wide spectral range $>$250 THz with a translation efficiency of (30.1~$\pm$~2.8)~\% and using an ultra-low pump power of (329~$\pm$~13)~$\mu$W. The translation efficiency projects to (274~$\pm$~28)~\% at 1~mW and is more than an order of magnitude larger than what has been achieved in current nanophotonic devices.

\end{abstract}


\maketitle
At its heart, nonlinear optics~\cite{Boyd2008,Agrawal2007} enables the creation of spectral tones at frequencies that are different than those at the input of the optical system.  As the name implies, such functionality is typically unavailable at low optical powers in solid-state systems, where most materials tend to exhibit only a linear optical response.  In addition, the spectral separation between the generated and input frequencies is often limited by the dispersive properties of common nonlinear media, which make phase-matching challenging. Nanophotonic systems, however, can realize large optical intensities for modest input powers, due to the strong field confinement and long photon lifetimes provided in resonator geometries.  The strong field confinement also creates geometric dispersion to balance material dispersion, so that phase-matching can be achieved for wide frequency separations. Here, we demonstrate a chip-scale system that links telecommunications-band photons and visible wavelength photons over a spectral separation $>~250$~THz, via ultra-low-power nonlinear nanophotonics, with a high translation efficiency achieved for sub-milliwatt pump power. The performance of our platform, quantified by translation efficiency relative to pump power, exceeds that of chip-integrated second-order nonlinear ($\chi^{(2)}$) systems, despite our use of the much weaker third-order nonlinear ($\chi^{(3)}$) effect. Our work illustrates the ability to use resonant enhancement and precise nanophotonic device engineering to dramatically enhance nonlinear optical processes.

Such spectral translation, i.e., the ability to bridge widely separated optical frequencies, has many applications. For example, high-performance tunable lasers have been developed in the context of heterogeneously-integrated silicon photonics~\cite{Komljenovic2016,Spencer2018}, but their operating wavelengths are restricted to $>1100$~nm due to the materials involved. Spectral translation to shorter wavelengths, e.g., the short near-infrared and visible, can make such sources suitable for spectroscopy of atomic systems, for applications such as wavelength references and optical clocks~\cite{Ludlow2015}. Moreover, the potential to link visible-wavelength systems together with telecom signals can be important in the context of long-distance fiber-optic distribution of timing and sychronization signals~\cite{riehle_optical_2017}.

\begin{figure*}[t]
\centering\includegraphics[width=0.7\linewidth]{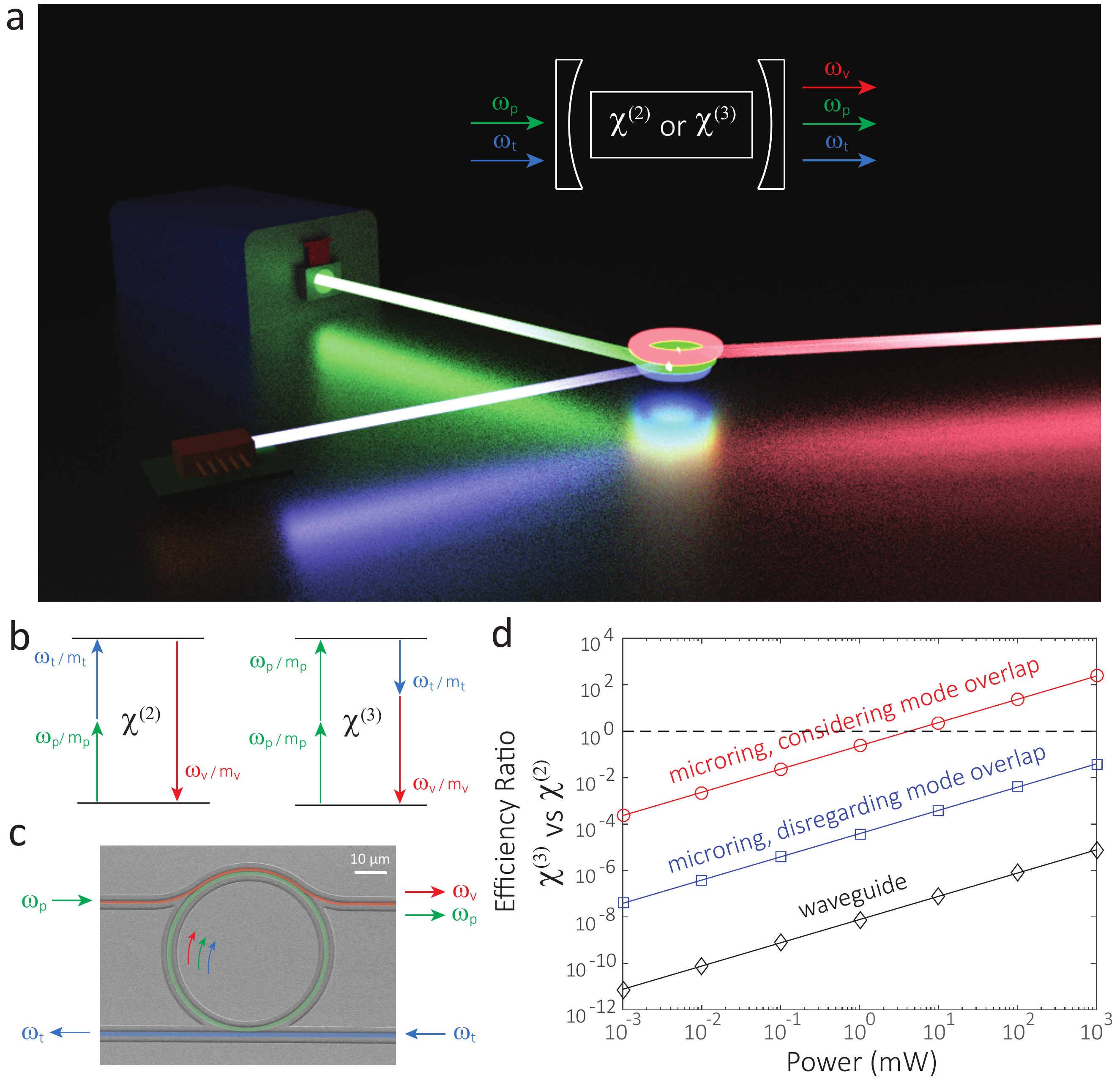}
\caption{{\bf Nanophotonic telecom-to-visible spectral translation and efficiency comparison.} \textbf{a}, Nanophotonic spectral translation uses a cavity-enhanced second-order or third-order nonlinear optical process ($\chi^{(2)}$ or $\chi^{(3)}$)  to efficiently transfer light into a new frequency with ultra-low laser pump power (see inset for a simplified scheme). $\omega_p$, $\omega_t$, and $\omega_v$ represent frequencies for pump, telecom, and visible light, respectively. \textbf{b}, Energy and momentum conservation requirements for sum-frequency generation (SFG, left) and degenerate four-wave mixing (dFWM, right) processes with single fundamental mode family (SFMF) operation. \textbf{c}, False-colored scanning-electron-microscope image of the nanophotonic device and the coupling scheme we use for spectral translation. The top pulley waveguide (red) and the bottom straight waveguide (blue) are used to couple pump/visible and telecom light, respectively, into and out of the microring (green). \textbf{d}, Order-of-magnitude comparison of the translation efficiency for the two processes in (b) with the device geometry of (c). For a microring with a finesse of $\mathcal{F}~\approx~5000$ (red), the efficiency of the $\chi^{(3)}$ process can compete with, or even exceed, that of the $\chi^{(2)}$ process at mW-pumping levels, if the mode overlap can be sufficiently well-optimized. See Supplementary Information for details.}

\label{Fig:Scheme}
\end{figure*}

\noindent \textbf{Nanophotonic spectral translation} Various forms of spectral translation have been realized using $\chi^{(2)}$ or $\chi^{(3)}$ in several different platforms, including lithium niobate\cite{Ilchenko2004,Furst2010,Wang2018,Luo2018}, silicon\cite{Liu2012}, silicon nitride \cite{Agha2013,Li2016,Ramelow2018,Hickstein2018,Grassani2019}, silicon dioxide\cite{Zhang2018}, aluminum nitride \cite{Guo2016,Guo2016a, Bruch2018}, gallium arsenide \cite{Chang2018,Chang2018a}, gallium phosphide \cite{Lake2016}, and aluminum gallium arsenide \cite{Mariani2014}. The $\chi^{(2)}$ nonlinear optical process has been the major workhorse for spectral translation across widely separated wavelengths for two main reasons. First, the magnitude of the nonlinear response resulting from the $\chi^{(2)}$ coefficient is much larger than that of higher-order nonlinearities. For example, in bulk nonlinear crystals, the laser intensity needs to be comparable to the strength of a characteristic atomic field ($I_{at}=3.5\times10^{20}~{\rm W/m^2}$) for the nonlinear response of a $\chi^{(3)}$ process to match that of a $\chi^{(2)}$ process\cite{Boyd2008}. Second, $\chi^{(2)}$ processes such as second-harmonic generation (SHG) and sum-frequency generation (SFG) naturally result in large spectral separations. On the other hand, $\chi^{(3)}$ nonlinear optics, within silicon nitride devices in particular, though having seen successful use in quantum frequency conversion \cite{Li2016}, entangled photon-pair generation \cite{Lu2018}, octave-span frequency comb generation \cite{Okawachi2011,Li2017,Karpov2018}, and optical frequency synthesis \cite{Spencer2018}, generally do not meet the performance levels of state-of-the-art $\chi^{(2)}$ spectral translation \cite{Furst2010}, in terms of the combination of efficiency, range, and required pump power.  For example, mm-scale LiNbO$_3$ resonators reach 9~$\%$ conversion efficiency at 30~$\mu$W pump power~\cite{Furst2010}, while in $\chi^{(3)}$ platforms, such a conversion efficiency has only been realized for wide spectral separations at much higher (tens of mW) pump power~\cite{Liu2012,Li2016}.

Nanophotonic spectral translation uses optical resonators to enhance the nonlinear optical processes (Fig.~\ref{Fig:Scheme}(a)), which enables ultra-low pump power operation at the expense of bandwidth, so that the primary application is translation of continuous-wave (CW) signals. We propose that, for telecom-to-visible spectral translation, high-$Q$ resonators with proper nanophotonic mode engineering can enable $\chi^{(3)}$ spectral translation to be comparable to the current state-of-the-art $\chi^{(2)}$ process. To illustrate this point, we compare the translation efficiency of $\chi^{(2)}$ and $\chi^{(3)}$ processes, e.g., SFG and degenerate four-wave mixing (dFWM) (Fig.~\ref{Fig:Scheme}(b)), in microring/waveguide geometries (Fig.~\ref{Fig:Scheme}(c)) using parameters for Si$_3$N$_4$ and LiNbO$_3$ as representative of $\chi^{(3)}$ and $\chi^{(2)}$ media, respectively (Fig.~\ref{Fig:Scheme}(d)). Translation efficiency is defined as the ratio of on-chip power between the visible translated signal and input telecom signal (see Methods for details). As described in the Supplementary Information, the ratio of the efficiency of the $\chi^{(3)}$ process to that of the $\chi^{(2)}$ process scales with pump power, so that a sufficiently large pump power can make these two processes comparable in efficiency. In the waveguide, the $\chi^{(3)}$ process is much weaker than $\chi^{(2)}$ for all powers considered, with a ratio of about -81~dB at 1~mW pump power (black). In a high-finesse microring ($\mathcal{F}~\approx~5000$), due to the cavity enhancement of the field intensity, the difference between these two processes becomes smaller, $\approx$-44~dB at 1~mW (blue). Mode matching is assumed to be ideal in the above estimates. However, in practice, demonstrated $\chi^{(2)}$ nonlinear resonators have typically exhibited non-ideal mode overlap~\cite{Ilchenko2004,Furst2010}, particularly in larger resonators that support many mode families so that mode identification is challenging.  In contrast, accurate mode identification has been recently demonstrated in ${\rm Si_3N_4}$ nanophotonics~\cite{Lu2018}, providing an opportunity for cavity-enhanced $\chi^{(3)}$ to exhibit an efficiency comparable to $\chi^{(2)}$ if an optimized mode overlap is realized. For example, the red curve in Fig.~\ref{Fig:Scheme}(d) indicates that an optimized $\chi^{(3)}$ process can be as efficient as current state-of-the-art $\chi^{(2)}$ at mW-level pump powers (red).

To achieve such an optimized $\chi^{(3)}$ spectral translation, three core elements are required for the the nonlinear optical resonator (Fig.~\ref{Fig:Scheme}(c)) and its mode engineering. First, simultaneous frequency-matching and phase-matching are required to satisfy energy conservation and momentum conservation (Fig.~\ref{Fig:Scheme}(b)). Second, high quality factor ($Q$) and strong modal confinement are required to create large circulating optical intensities. Third, large spatial overlap between the modes is critical to maximize the interaction efficiency. We use cavity-enhanced degenerate four-wave mixing (dFWM) in a silicon nitride microring resonator (Fig.~\ref{Fig:Scheme}(c)) to translate a continuous-wave signal at 1573~nm to a visible wavelength at 670~nm (a spectral range over 250~THz), with a translation efficiency of (30.1~$\pm$~2.8)~\% at only (329~$\pm$~13)~$\mu$W pump power. This value projects to (274~$\pm$~28)~\% at 1~mW and is at least one order of magnitude higher than that reported from other nanophotonic devices, and is comparable to the best result using a $\chi^{(2)}$ process, achieved using mm-scale resonators~\cite{Furst2010}.

\noindent \textbf{System design}
The previous theoretical analysis shows that both optical quality factor and nanophotonic mode engineering play critical roles for the efficiency of the $\chi^{(3)}$ spectral translation. Our design target here is therefore a mode overlap of $\approx100~\%$ for the dFWM process and a large $Q/V$ for all three interacting modes (in the telecom, pump, and visible bands). As a practical matter, we prefer geometries that support a limited number of mode families (particularly in the visible), to  ease the task of identifying and employing the targeted optical modes. Moreover, efficiently coupling the telecom, pump, and visible modes of the microring to the access waveguide is also necessary, both for coupling the pump/signal light into the microring and extracting the translated light out of the microring. We outline the basic design principles in this section.

The dispersion engineering required for a microring differs from that for a straight waveguide, in which phase-matching means that linear momentum of the input/output photons must be conserved. In a microring, phase-matching applies to the momentum of the whispering gallery modes along the azimuthal direction. We define the whispering gallery modes with the same polarization and radial mode order to be within one mode family. Within each mode family, a whispering gallery mode is specified by its azimuthal number $m$ (electric field varies along the azimuthal angle ${\phi}$ as ${e}^{i m \phi}$. The phase-matching condition for modes of the same family becomes a mode number matching criterion, i.e., ${\Delta} m = 0$. For dFWM, the criterion is $2m_\text{p}-m_\text{t}-m_\text{v}=0$, where $m_\text{p}/m_\text{t}/m_\text{v}$ are the azimuthal mode numbers for the pump/telecom (input signal)/visible (translated signal) modes \cite{Kippenberg2004}.

Frequency matching in a microring is more challenging than in a waveguide due to the high-$Q$ cavity resonances. Once phase-matching is satisfied for an appropriate set of modes, the energy conservation condition requires ${2\omega_\text{p}-\omega_\text{t}-\omega_\text{v}=0}$, and the involved pump/telecom/visible fields (${\omega_\text{p}/\omega_\text{t}/\omega_\text{v}}$) all have to be in resonance with the corresponding cavity modes (${\hat{\omega}_\text{p}/\hat{\omega}_\text{t}/\hat{\omega}_\text{v}}$). The maximum allowable frequency mismatch for the dFWM process is dictated by the loaded $Q$ of the corresponding cavity modes, i.e., ${\delta\omega_\text{p,t,v} = |\omega_\text{p,t,v} - \hat{\omega}_\text{p,t,v}|~<~\hat{\omega}_\text{p,t,v}/2Q_\text{p,t,v}}$. This stringent frequency-matching requirement represents a major challenge for nanophotonic devices, necessitating the matching of frequencies separated over hundreds of THz with an accuracy within a cavity linewidth, e.g., $\approx$~1~GHz for a loaded $Q$ of $2\times10^5$ at a telecom wavelength.

Our microring geometry uses a stochiometric silicon nitride (Si$_3$N$_4$) core, bottom silicon dioxide cladding, and a top air cladding. This has two main benefits. First, it allows post-processing for dispersion tuning. For example, a post-fabrication reactive-ion etching process can reduce the device height while maintaining the device lateral dimensions. Also, diluted hydrofluoric acid etching can change the device height and lateral dimensions simultaneously with nanometer accuracy. These two post-processing methods, when used with ellipsometric measurement, enable fine tuning of the dispersion of targeted mode sets for fabricated devices. Second, as described below, the asymmetric cladding allows for separation of the coupling tasks for pump/visible and telecom modes in a manner that would be difficult to implement in symetrically oxide-clad devices, by taking advantage of waveguide cutoff~\cite{Yariv2006}.

\begin{figure}[t]
\centering\includegraphics[width=1\columnwidth]{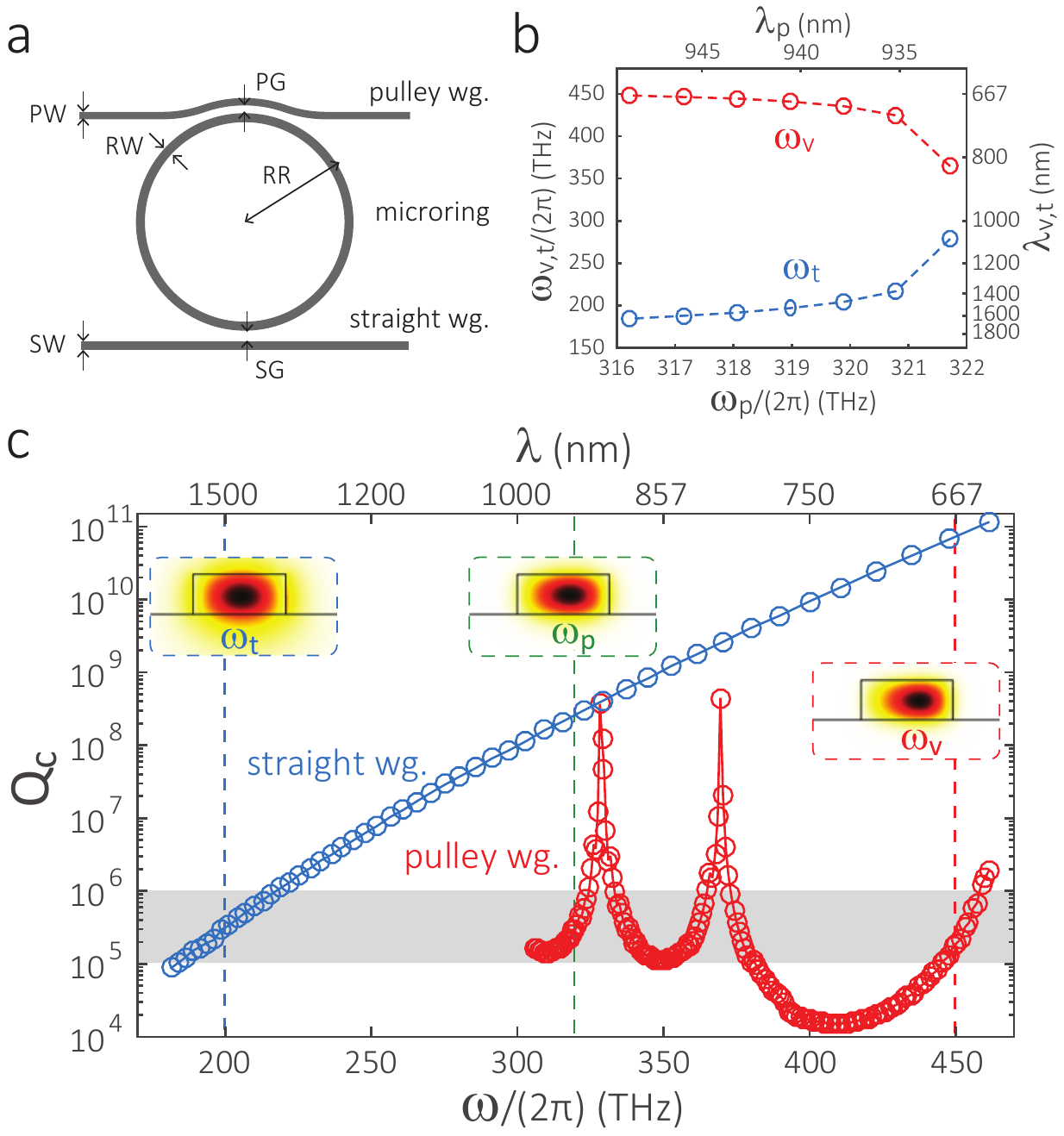}
\caption{{\bf Design for telecom-to-visible spectral translation.} \textbf{a}, Device scheme. $RR$/$RW$: ring outer radius/ring width. $PW$/$PG$: pulley waveguide width/gap. $SW$/$SG$: straight waveguide width/gap. $wg$: waveguide. \textbf{b}, Simulations of the frequency/phase-matched wavelengths. The $m$ numbers of the three modes satisfy (${2m_\text{p} = m_\text{t} + m_\text{v}}$), with a frequency mismatch ${\Delta\omega/(2\pi)=|2\omega_\text{p} - \omega_\text{t}-\omega_\text{v}|/(2\pi)}$ within 1~GHz. Simulation parameters are $H$ = 500~nm, $RR$ = 25~$\mu$m, and $RW$ = 1158~nm. \textbf{c}, Wavelength-dependent coupling of the pulley (red) and straight (blue) waveguides. Targeted values are typically $Q_\text{c}=10^5$ to $10^6$ (gray area). All modes are fundamental transverse electric modes (TE1), with a cross-section view of the dominant electric field amplitude (in the radial direction of the microring) shown in the insets. Simulation parameters (in addition to \textbf{b}): $PW$ = 560~nm, $PG$ = 170~nm, $SW$ = 1120~nm, and $SG$ = 425~nm.}
\label{Fig:Simulation}
\end{figure}

\noindent \textbf{Single fundamental mode family engineering}
The heart of our design is single fundamental mode family (SFMF) engineering, where all three interacting modes are in the fundamental transverse-electric mode family (TE1), with the dominant electric field component in the radial direction (insets of Fig.~\ref{Fig:Simulation}(c)). The benefits of SFMF engineering are twofold. First, fundamental modes typically have the highest $Q$s, maximizing the resulting optical intensities (for a given pump power) that are critical for the efficiency of the spectral translation process.  Second, SFMF engineering can achieve a mode overlap of $\approx$~90~\% for dFWM across telecom, pump, and visible bands.  We note that, while the use of fundamental modes for nonlinear processes is common within the telecommunications band, e.g., in spontaneous four-wave-mixing devices~\cite{Caspani2017}, for operation over widely separated frequencies, higher order modes are often used. For example, $\chi^{(2)}$ nanophotonic devices used in SHG generally involve a higher order mode family for the SHG mode. In addition to lower $Q$ factors, their overlap with fundamental modes is far from ideal - it is negligible between TE1 and TE2, and $\approx$~0.3 for TE1 and TE3 in our geometry.

To demonstrate the potential range of $\chi^{(3)}$ spectral translation, we target a spectral translation window that is over one octave (i.e., comparable to SFG), in translating a telecom wavelength of 1570 nm to a wavelength of 670 nm by a pump located at 940 nm. A schematic of the nanophotonic spectral translation device is shown in Fig.~\ref{Fig:Simulation}(a). The microring is defined by three parameters, its height ($H$), outer ring radius ($RR$), and ring width ($RW$). These parameters yield geometric dispersion (combination of waveguiding and bending effects) that compensates material dispersion across the three bands of the dFWM process. We choose $H=$~500~nm, $RR=~$25~$\mu$m, and $RW=~$1150~nm to target the selected dFWM wavelengths. A finite-element-method simulation is carried out to determine the frequency- and phase-matched mode sets for the TE1 family (Fig.~\ref{Fig:Simulation}(b)). For a pump wavelength around 940 nm, we have frequency- and phase-matched telecom and visible modes near 1570~nm and 670~nm, respectively. The frequency matching criterion used in this simulation is ${\Delta\omega/(2 \pi)<1}$~GHz.

The wide spectral separation of the dFWM process requires careful waveguide-microring coupling design. We use two waveguides to separate the coupling tasks, as shown in Fig.~\ref{Fig:Simulation}(a). The top pulley waveguide\cite{Li2016} is wrapped around the microring to provide a 33~$\mu$m interaction length that couples the 940~nm and 670~nm modes efficiently. This waveguide has a width of 560~nm and supports single mode operation at 940~nm, while being cut-off (i.e., does not support any mode) at 1550~nm. The bottom straight waveguide has a width of 1120~nm and is single mode at 1550~nm. The wavelength-dependent coupling $Q$ factors ($Q_{c}$) are calculated using coupled mode theory~\cite{Hosseini2010, Li2016}, and shown in Fig.~\ref{Fig:Simulation}(c) for both straight waveguide (blue) and pulley waveguide (red). The targeted $Q_{c}$ is $10^5$ to $10^6$, which is indicated by the grey area. The straight waveguide couples the 1570~nm mode efficiently, but essentially does not couple the 940~nm and 670~nm modes at all (severe undercoupling due to the large coupling gap). The pulley waveguide, while cut-off at telecom wavelengths, couples the 940~nm and 670~nm modes efficiently.

\begin{figure}[t]
\centering\includegraphics[width=1.0\columnwidth]{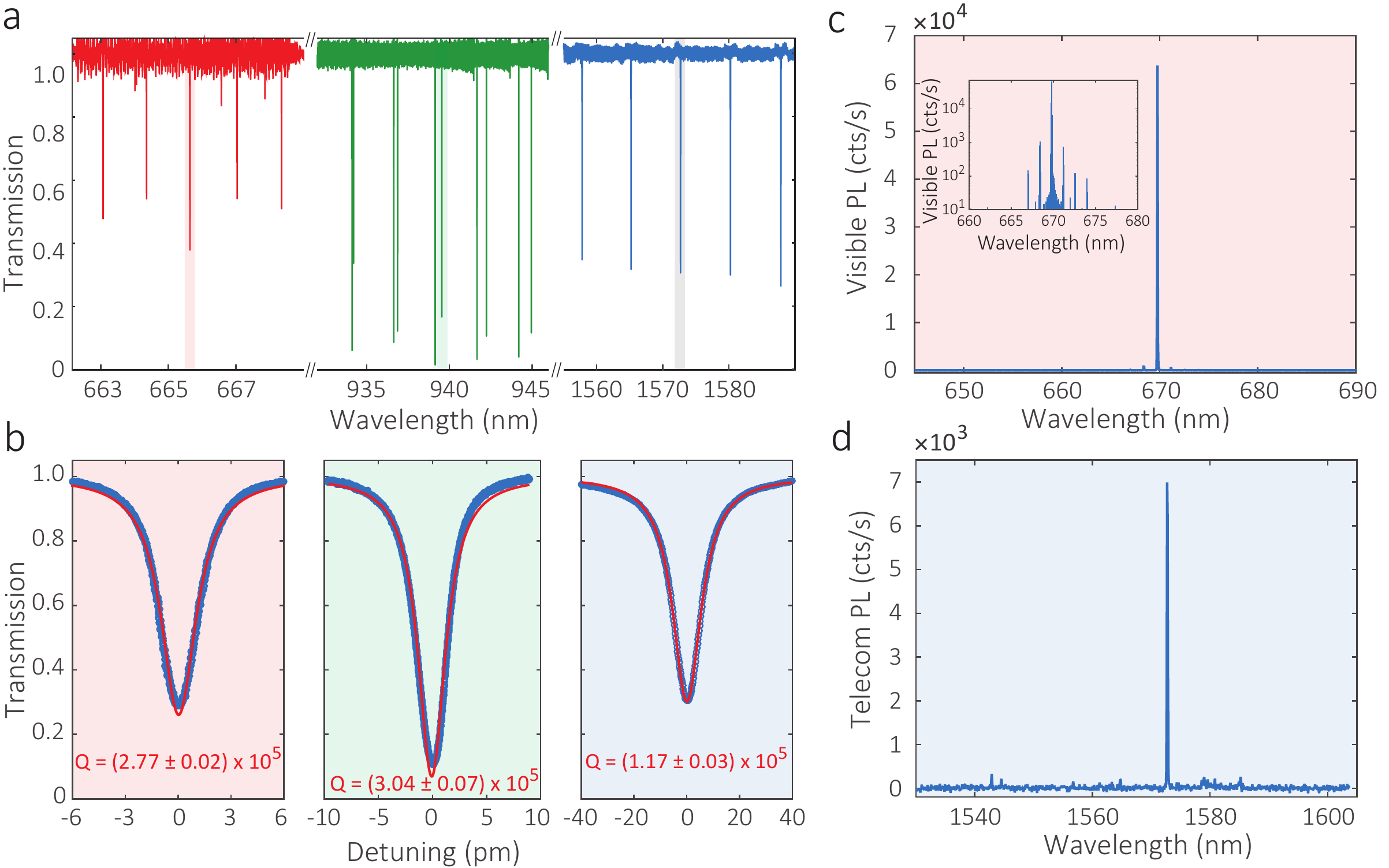}
\caption{{\bf Assessment of device $Q$, coupling, and phase- and frequency-matching.} \textbf{a}, Cavity transmission for the spectral translation device in visible, pump, and telecom bands, with device parameters prescribed in Fig.~\ref{Fig:Simulation}(c). \textbf{b}, Zoom-in transmission traces for TE1 modes at 665.8~nm (red), 939.5~nm (green), and 1572.7~nm (blue) from left to right with loaded $Q$ factors of $\approx~(1-3)\times10^5$ estimated by Lorentzian fitting (red lines). \textbf{c-d}, Visible-telecom photon-pair spectra by spontaneous four-wave mixing, with a degenerate pump at 939.5~nm. Inset of \textbf{\c} shows the visible spectrum in log scale. The 669.8~nm/1572.7~nm photon-pair spectra are free from broadband noise and are $>$~16~dB larger than adjacent mode sets, which indicates good frequency- and phase-matching for the dFWM process.}
\label{Fig:Trans}
\end{figure}

\noindent \textbf{Device characterization}
The device design described in the previous section is fabricated (see Methods) and characterized experimentally (see Supplementary Information for the experimental setup). The cavity linear transmission spectrum is shown in Fig.~\ref{Fig:Trans}(a), where the traces indicated by different colors are measured using separate lasers. For TE polarization, the transmission spectrum shows a single mode family in the telecom band and only two mode families that are relatively efficiently coupled in the pump and visible bands. The TE1 coupling is indeed reasonably efficient as designed, with 670~nm, 940~nm, and 1550~nm modes overcoupled, nearly critically-coupled, and slightly undercoupled, respectively. The intrinsic $Q$s are $(1.13\pm0.01)\times10^6$, $(8.32\pm0.19)\times10^5$, and $(5.29\pm0.14)\times10^5$, respectively, for the visible, pump, and telecom modes. The loaded $Q$s are $(2.77\pm0.02)\times10^5$, $(3.04\pm0.07)\times10^5$, and $(1.17\pm0.03)\times10^5$, respectively, with the transmission traces and fits shown in Fig.~\ref{Fig:Trans}(b). The uncertainties are one standard deviation value resulting from the Lorentzian fits. The candidate set of mode numbers for dFWM is 442, 301, and 160, and is identified using a mode splitting method that targets nearby azimuthal modes~\cite{Lu2014, Lu2018}.

As an approach to evaluate whether the chosen set of phase-matched modes is indeed frequency-matched, we use spontaneous photon pair generation. When phase-matching and frequency-matching are satisfied, both spontaneous and stimulated degenerate four-wave mixing can occur. We can thus use the spontaneous photon-pair spectra to evaluate the phase- and frequency-matching for the stimulated process in the regime for which nonlinear frequency shifts are small~\cite{Helt2012}. We record the spontaneous photon-pair spectra with a 939.5~nm pump and without any narrow-band filters in Fig.~\ref{Fig:Trans}(c)-(d). The brightest set of spectral tones are at 669.8~nm/1572.7~nm, and the adjacent modes are $>$16~dB smaller (inset of (c)). The photon-pair spectrum suggests that good phase- and frequency-matching has been achieved for the aforementioned mode set.

\begin{figure*}[t]
\centering\includegraphics[width=0.8\linewidth]{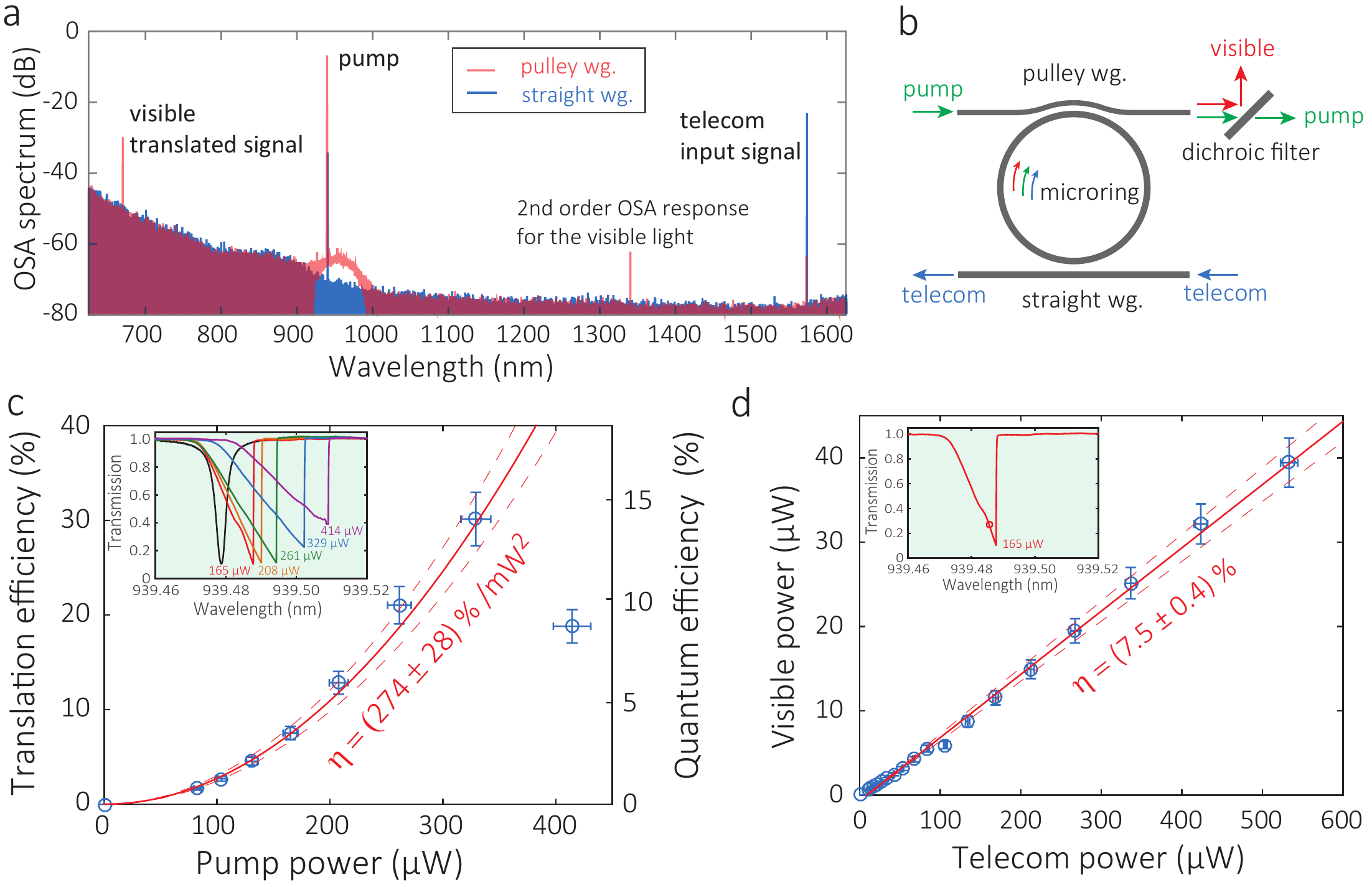}
\caption{{\bf Telecom-to-visible nanophotonic spectral translation.} \textbf{a}, Optical spectra recorded in the pulley (red) and straight (blue) waveguides for the spectral translation device. The telecom light at 1572.7~nm (blue) is transferred to a visible wavelength at 669.8~nm (red) through a pump at 939.5~nm, with no other translation channels or noise contribution observed. 0 dB is referenced to 1 mW (i.e., dBm). \textbf{b}, A dichroic filter is used to reject the pump light ($>$ 80 dB) in order to calibrate the visible power accurately in pump-power dependent measurements. \textbf{c}, Translation efficiency ($\eta$, left y-axis) and quantum efficiency ($\eta_\text{Q}$, right y-axis) versus pump power. A translation efficiency $\eta~=~$(30.1~$\pm$~2.8)~\% is achieved for (329~$\pm$~13)~$\mu$W pump power. The quadratic dependence on pump power is a signature of the degenerate four-wave mixing process. Solid and dashed red lines represent the quadratic fitting and its one standard deviation confidence range. Insets show the corresponding pump transmission traces. The pump detuning is brought close to the bottom of the transmission dip to maximize the translation efficiency. \textbf{d}, Output visible power as a function of input telecom power for a fixed pump power of (165~$\pm$~7)~$\mu$W, with a translation efficiency determined by a linear fit of $\eta~=~$(7.5~$\pm$~0.4)~\%. The pump detuning is kept constant as indicated by the open circle in the inset transmission trace. Solid and dashed red lines represent the linear fitting and its one standard deviation confidence range. Error bars in \textbf{c-d} are one standard deviation uncertainties originating from calibration of the on-chip power.}
\label{Fig:SpecTrans}
\end{figure*}

\noindent \textbf{Spectral translation}
Spectral translation is then initiated by sending to the device an input telecom signal at 1572.7~nm in addition to the pump laser at 939.5~nm. The measured optical spectra from the pulley and straight waveguide outputs are overlaid in Fig.~\ref{Fig:SpecTrans}(a). These spectra are taken directly from the waveguides using lensed single-mode optical fibers without any filters, and confirm that the stimulated process translates the 1572.7~nm signal to 669.8 nm. We observe that the pulley waveguide (red) out-couples pump and visible light, while the straight waveguide (blue) out-couples telecom light. The isolation at 1572.7 nm in the pulley waveguide is about 40~dB, consistent with the waveguide cutoff previously described, while the straight waveguide, as expected, does not show appreciable out-coupling of the visible translated signal. The spectra are relatively clean, with no broadband or narrowband noise or competing four-wave-mixing tones observed (the small peak near 1340 nm is an artefact of the optical spectrum analyzer).

We optimize the translation efficiency at various pump/signal power levels and show the power-dependence in Fig.~\ref{Fig:SpecTrans}(c)-(d). To measure the translated power accurately, we use a dichroic filter to separate the 939.5~nm pump and 669.8~nm translated signal (Fig.~\ref{Fig:SpecTrans}(b); see Supplementary Information for details). The pump power dependence of the telecom-to-visible translation efficiency is plotted in Fig.~\ref{Fig:SpecTrans}(c), where the observed quadratic behavior (red lines) is a signature of the degenerate pump in the dFWM process. Two types of efficiency, i.e., translation efficiency ($\eta$) and quantum efficiency ($\eta_\text{Q}$), are plotted. As defined earlier, the translation efficiency is based on a ratio of powers and is consistent with what is often quoted in classical applications (such as SHG), while quantum efficiency is defined in terms of photon flux (See Methods for the precise definitions). A peak translation efficiency $\eta=(30.1~\pm~2.8)~\%$ is reached, corresponding to a quantum efficiency $\eta_\text{Q}=(12.8~\pm~1.2)~\%$. At the highest measured pump power of (414~$\pm$~17)~$\mu$W, the translation efficiency drops due to close-to-pump parametric oscillation and cascaded four-wave mixing and its consumption of pump power. The consumption of pump power can be partly verified by the reduction in resonance contrast in the pump transmission traces (inset to Fig.~\ref{Fig:SpecTrans}(c)). This issue can potentially be addressed in the future by advanced dispersion or coupling designs to suppress the competing process.

The translated signal power at 669.8~nm has a linear dependence on the input power at 1572.7~nm, as shown in Fig.~\ref{Fig:SpecTrans}(d) for an intermediate pump power of (165~$\pm$~7)~$\mu$W. The linear fit (red line) yields a translation efficiency $\eta~=~$(7.5~$\pm$~0.4)~\%. The efficiency is taken with pump detuning kept at a constant level indicated by the inset of Fig.~\ref{Fig:SpecTrans}(d). The maximum signal power here is (533~$\pm$~11)~$\mu$W, which is larger than the pump power of (165~$\pm$~7)~$\mu$W, while the translation efficiency is still not saturated. This is possible because the pump photon number/energy in the microring is sufficient, as a result of higher optical quality factor and better waveguide-microring coupling for the 940~nm mode. An even higher telecom power leads to pump depletion and loss of the fixed pump detuning due to the thermal bistability of the cavity mode. For applications that need higher powers, the pump power can be increased as shown in Fig.~\ref{Fig:SpecTrans}(c), up to the point where competing four-wave mixing processes ultimately limit the pump power available for spectral translation.

\noindent \textbf{Comparison}
To evaluate the performance of our nanophotonic device among state-of-the-art wide-band spectral translation devices, we compare it with reported results from the literature in Table~\ref{Tab:Comparison}. In this table we consider material platform, device geometry, nonlinear process, dimensions, operation wavelengths, pump power, demonstrated efficiency, and expected efficiency at 1~mW pump power. Our device has the smallest footprint among the nanophotonic devices with on-chip integrated waveguides and is 50 to 1600 times smaller than the mm/cm devices. The spectral translation range is $>$~250~THz and more than an octave, and could be further increased or shifted to other useful spectral bands by simply changing the pump modes (Fig.~\ref{Fig:Simulation}(b)) and/or reducing the device height and ring width.  The typical operating power is sub-mW, which is on par or lower than most other high-$Q$ nanophotonic devices, particularly in light of our high translation efficiency of 30~\%. While Ref.~\citen{Li2016} achieves a higher translation efficiency, it uses over 100 times more pump power. To compare with other results in terms of translation efficiency, we consider the measured (or expected) efficiency at 1~mW pump power. Our device has the highest efficiency of (274~$\pm$~28)~\% at 1~mW among nanophotonic devices, by at least one order of magnitude. This normalized efficiency is also larger than most of the mm/cm devices, and is very similar to the highest value in a $\chi^{(2)}$ resonator, reported in Ref.~\citen{Furst2010}.
\begin{table*}
\begin{center}
\scalebox{0.80}{
       \begin{tabular}{ c c | c c c c c c c | c}
        \hline
        \hline
             Label & Reference & Material & Geometry  &  Nonlinear process  & Dimensions~($\mu$m) & $\lambda_s$-$\lambda_t$~(nm)  &    P~(mW)       & $\eta$(P)  &  $\eta$(P=1~mW)\\ [0.5ex]
        \hline
            \scalebox{1.5}{${\color{red}{\bullet}}$}~&~this work & ${\rm Si_3N_4}$ & microring & $\chi^{(3)}$, dFWM  & 25/1.15/0.5 & 1570-670 & 0.329 & 30.1~\% & 274~\% \\
            \scalebox{1.0}{${\color{blue}{\otimes}}$}~&~$[$\citen{Guo2016}$]$ & AlN &  microring & $\chi^{(2)}$, SFG  & 30/1.12/1  & 1538-774 & 9.8 & 25~\%$^{c}$ &  2.6~\%$^{c}$\\
            &~$[$\citen{Guo2016a}$]$ & AlN &  microring & $\chi^{(2)}$, SHG  & 30/1.12/1  & 1544-772 & 27 & 12~\% &  2.5~\%\\
            \scalebox{1.0}{${\color{blue}{\diamondsuit}}$}~&~$[$\citen{Luo2018}$]$ & LiNbO$_3$ &  microring & $\chi^{(2)}$, SHG  & 50/0.69/0.55  & 1550-775 & 0.44 & 0.66~\% &  1.5~\%\\
               &~$[$\citen{Bruch2018}$]$ & AlN &  microring & $\chi^{(2)}$, SHG  & 30/1.2/1  & 1560-780 & 0.96 & 1.2~\% &  1.3~\%$^c$/17~\%$^i$\\
            \scalebox{1.5}{${\color{red}{\circ}}$}~&~$[$\citen{Li2016}$]$ & ${\rm Si_3N_4}$ &  microring & $\chi^{(3)}$, FWM-BS & 40/1.4/0.48 & 1530-960  & 50/8  & 96~\%$^c$ &  0.24~\%$^c$ \\
              &~$[$\citen{Lin2016}$]$ & ${\rm LiNbO_3}$ &  microdisk & $\chi^{(2)}$, SHG & 51/0.7 & 1540-770 & 10 & 1.1~\% & 0.11~\% \\
            \scalebox{1.0}{${\color{blue}{\triangle}}$}~&~$[$\citen{Chang2018}$]$  & GaAs &  microring NA &  $\chi^{(2)}$, SHG & 100/1.3/0.15  & 2000-1000  &    5   & 0.2~\%  & 0.04~\%$^c$/65~\%$^i$\\
            \scalebox{1.5}{${\color{blue}{\triangleleft}}$}~&~$[$\citen{Lake2016}$]$ & ${\rm GaP}$ &  microdisk & $\chi^{(2)}$, SHG & 3.3/0.25 & 1545-772  & 0.35 & $(1.54\times10^{-4})~\%^c$ &  $(4.4\times10^{-4})~\%$ \\
            \scalebox{1.5}{${\color{blue}{\triangleright}}$}~&~$[$\citen{Levy2011}$]$ & ${\rm Si_3N_4}$ &  $\mu$-ring & $\chi^{(2)}$, SHG & 116/1.5/0.725 & 1554-777  & 64  & 0.013~\% &  $(2.0\times10^{-4})~\%^c$ \\
            \scalebox{1.0}{${\color{red}{\diamondsuit}}$}~&~$[$\citen{Surya2017}$]$ & AlN/${\rm Si_3N_4}$ &  microring & $\chi^{(3)}$, THG  & 30/1.2/1  & 1560-780 & 30 & 0.16~\% &  $(1.8\times10^{-4})~\%$\\
            \scalebox{1.3}{${\color{blue}{\triangledown}}$}~&~$[$\citen{Mariani2014}$]$ & AlGaAs &  microdisk & $\chi^{(2)}$, SHG  & 1.9/0.155  & 1584-792 & 1.0 & $(7.0\times10^{-5})~\%$ &  $(7.0\times10^{-5})~\%$\\
            \scalebox{1.0}{${\color{blue}{\Box}}$}~&~$[$\citen{Zhang2018}$]$ & ${\rm SiO_2}$ &  microsphere & $\chi^{(2)}$, SHG & 62  & 1550-775 & 0.88$^{c}$ & ${(4.3\times10^{-5})~\%}^c$ & ${(4.9\times10^{-5})~\%}$\\

        \hline
            \scalebox{1.5}{${\color{blue}{\bullet}}$}~&~$[$\citen{Furst2010}$]$ & ${\rm LiNbO_3}$ &  mm-resonator & $\chi^{(2)}$, SHG & 1900/500 & {1064-532}  &   0.03 & 9~\% & ${300~\%}$\\
            \scalebox{1.5}{${\color{blue}{\circ}}$}~&~$[$\citen{Ilchenko2004}$]$ & PPLN &  mm-resonator & $\chi^{(2)}$, SHG & 1500/500 & {1550-775} & 9 & 22~\% & 2.4~\%$^{c}$ \\
            \scalebox{1.0}{${\color{blue}{\times}}$}~&~$[$\citen{Wang2018}$]$ & PPLN &  waveguide &  $\chi^{(2)}$, SHG & 4000/1.4/0.6 & 1500-775  &    20     & 8~\%   & 0.4~\%$^{c,l}$ \\
            \scalebox{1.3}{${\color{blue}{*}}$}~&~$[$\citen{Chang2018a}$]$ & GaAs  & waveguide &  $\chi^{(2)}$, SHG & 1400 & 2000-1000  &    2   & 0.5~\%$^c$  & 0.25~\%$^{c,l}$ \\
            \scalebox{1.3}{${\color{red}{*}}$}~&~$[$\citen{Liu2012}$]$ & Si &  waveguide &  $\chi^{(3)}$, dFWM & 20000 & 3550-1590  & 300 & ${16000~\%}$ &   0.18~\%$^{c}$ \\
            \scalebox{1.0}{${\color{blue}{+}}$}~&~$[$\citen{Grassani2019}$]$ & ${\rm Si_3N_4}$ & waveguide & $\chi^{(2)}$, SFG & 40000/1.5/0.87 & 1547-773 & 91 & 0.031\% & $(3.3\times10^{-4})\%^{c}$\\
            \scalebox{1.0}{${\color{red}{\times}}$}~&~$[$\citen{Agha2013}$]$ & ${\rm Si_3N_4}$ & waveguide & $\chi^{(3)}$, FWM-BS & 18000/1.2/0.55 & 1500-980 & 11 & $(1.2\times10^{-5})~\%$ & $(9.9\times10^{-8})\%^{c}$\\
            \scalebox{1.0}{${\color{red}{+}}$}~&~$[$\citen{Grassani2019}$]$ & ${\rm Si_3N_4}$ & waveguide & $\chi^{(3)}$, dFWM & 40000/1.5/0.87 & 1547-1540 & 90 & $(7.9\times10^{-4})\%$ & $(9.8\times10^{-8})\%^{c}$\\
        \hline
      \end{tabular}
}
        \caption{\textbf{Comparison with related state-of-the-art spectral translation devices.} The table compares our device with other state-of-the-art devices, in terms of material, device geometry, $\chi^{(2)}$ or $\chi^{(3)}$ nonlinear process, dimensions, representative operating wavelengths, pump power, and efficiency. The table is organized in two sections, summarizing results from nanophotonic spectral translation devices (top) and mm/cm-scale photonic devices (bottom). Dimensions are specified as radius/width/height for microrings, radius/height for microdisks and mm-resonators, radius for microspheres, and length/width/height for waveguides. $\lambda_s$: signal wavelength, $\lambda_t$: translated wavelength, $\eta_P$: translation efficiency, dFWM: degenerate four-wave mixing, SHG: second-harmonic generation, SFG: sum-frequency generation, FWM-BS: four-wave mixing Bragg scattering. PPLN: periodically poled lithium niobate with poling periods of 14~$\mu$m and 4~$\mu$m in Ref.~\onlinecite{Ilchenko2004} and Ref.~\onlinecite{Wang2018}. $^c$: values calculated from representative data. $^i$: internal efficiency. $^l$: efficiency normalized by waveguide lengths is 2.6~\%/mW/${\rm cm^2}$ and 13~\%/mW/${\rm cm^2}$ in Ref.~\onlinecite{Wang2018} and Ref.~\onlinecite{Chang2018a}.}
    \label{Tab:Comparison}
\end{center}
\end{table*}

\begin{figure*}[t!]
\centering\includegraphics[width=1.5\columnwidth]{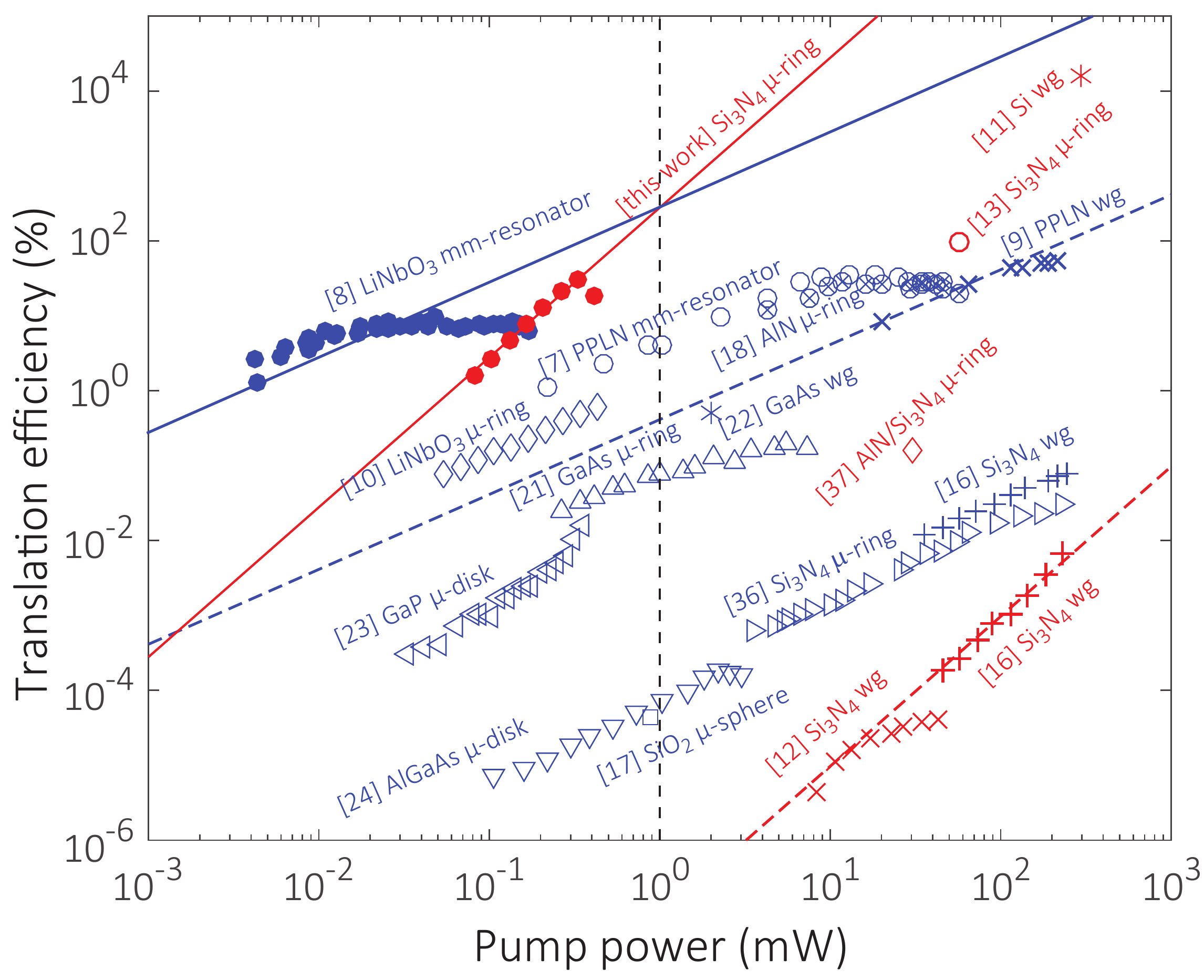}
\caption{{\bf Comparison of our nanophotonic spectral translation efficiency with state-of-the-art results.} We compare our results with other state-of-the-art devices in terms of translation efficiency vs. pump power. Solid/dashed lines show the scaling in $\chi^{(2)}$ ${\rm LiNbO_3}$ (blue) and $\chi^{(3)}$ ${\rm Si_3N_4}$ (red) resonators/waveguides, respectively, which agree with theoretical estimates in that $\chi^{(2)}$ and $\chi^{(3)}$ processes have linear and quadratic power scaling, respectively (Ref.~\onlinecite{Lake2016} is distinguished from others due to its cavity frequency mismatch.). Through resonance enhancement and careful mode engineering, our $\chi^{(3)}$ device is projected to reach a translation efficiency of (274~$\pm$~28)~\% at 1 mW, which is comparable to that of the best $\chi^{(2)}$ device.  This represents a many order of magnitude improvement relative to the waveguide case, where the $\chi^{(3)}$ process is over 50 dB smaller in efficiency than the $\chi^{(2)}$ case at 10 mW, $\mu$: micro. wg: waveguide.}
\label{Fig:Comparison}
\end{figure*}

Figure~\ref{Fig:Comparison} provides more detail in comparing the translation efficiency of our work to related studies. $\chi^{(3)}$ works (red symbols) have quadratic power dependence (solid/dashed red lines) and $\chi^{(2)}$ works (blue symbols) have linear power dependence (solid/dashed blue lines). For $\chi^{(2)}$ processes, we have considered both SHG and SFG. We note that although the SHG output signal has a quadratic dependence on pump power, the power scaling for translation efficiency is still linear (because the pump also serves as the input signal). Likewise, for non-degenerate FWM, the power scaling for translation efficiency is the same as degenerate FWM (see Supplementary Information for details). The solid/dashed lines highlight resonator/waveguide performance for ${\rm LiNbO_3}$ (blue) and ${\rm Si_3N_4}$ (red), respectively. The trends are similar to theoretical estimates (Fig.~\ref{Fig:Scheme}(d)). In the waveguide case, the $\chi^{(2)}$ process is much more efficient than the $\chi^{(3)}$ process ($\approx$55~dB difference at 10~mW pump power), because of the much larger nonlinear optical coefficient. However, in the resonator case (Ref.~\citen{Furst2010} and this work), the two processes are actually expected to be comparable to each other at 1~mW pump power. At 10~mW, our $\chi^{(3)}$ device is projected to be about 10 times more efficient than that projected for the ${\rm LiNbO_3}$ mm-resonator, provided that the aforementioned close-to-pump four-wave mixing could be suppressed. Our work provides a clear demonstration of how a $\chi^{(3)}$ nanophotonic spectral translation device can be comparable to, or even better than, the record $\chi^{(2)}$ device in terms of translation efficiency. We use 1~mW instead of 1~W (the commonly specified power level for SHG) for the comparison point since this is a reasonable operation power for high-Q photonic devices. A comparison at 1~W would be additionally 30~dB more advantageous for the $\chi^{(3)}$ process (Fig.~\ref{Fig:Comparison}).


Recently, Ramelow et al.~\cite{Ramelow2018} have discussed how one can view dFWM as an effective
$\chi^{(2)}$ process.  In our case, one considers the degenerate input (the pump) as being converted to the visible, with the other input (the telecom input in our case) as the pump. The power scaling of such a process is similar to SHG, and single-photon-level nonlinear interactions are in principle possible if the resonant enhancement is sufficiently strong. In Ref.~\citen{Ramelow2018}, the efficiency of conversion from a 350~$\mu$W signal at 1410~nm (the degenerate input) to 1250~nm is 5~$\%$, using an 85~mW pump at 1590~nm. To evaluate our work in a similar way, we treat the 1573~nm signal as the effective pump, the 940~nm light (the degenerate input) as the seed, and the 670~nm light as the converted output. Our efficiency (Fig.~\ref{Fig:SpecTrans}(d)) is (24~$\pm$~2)~\% at (165~$\pm$~7)~$\mu$W seed power, with an effective pump power of (533~$\pm$~11)~$\mu$W. In comparison to Ref.~\onlinecite{Ramelow2018}, this corresponds to a higher efficiency ($\approx~5\times$), a smaller pump power ($\approx~200\times$), and a wider translation range ($\approx~5\times$, or $\approx~$100~THz larger).

\noindent \textbf{Discussion}
In summary, by using single fundamental mode familty (SFMF) engineering and precise mode identification to enable phase- and frequency-matched operation, we have demonstrated high-efficiency spectral translation from telecom to visible wavelengths using degenerate four-wave mixing in a silicon nitride nanophotonic resonator. We propose and demonstrate that, for the first time, a $\chi^{(3)}$ device can be comparable to, or even superior than, state-of-the-art $\chi^{(2)}$ devices for wide-band spectral translation. The demonstrated translation efficiency at sub-mW pump power improves the current record in nanophotonic devices by at least one order of magnitude. This approach holds great promise for many applications, such as connecting visible wavelength atomic references to the telecommunications band.


\noindent \textbf{Methods}
\\
\noindent \textbf{Device Fabrication}
The device layout was done with the Nanolithography Toolbox, a free software package developed by the NIST Center for Nanoscale Science and Technology~\cite{coimbatore_balram_nanolithography_2016}. The 500~nm thick ${\rm Si_3N_4}$ layer is deposited by low-pressure chemical vapor deposition on top of a 3~${\rm \mu}$m thick thermal ${\rm SiO_2}$ layer on a 100~mm diameter Si wafer. The wavelength-dependent refractive index and the thickness of the layers are measured using a spectroscopic ellipsometer, with the data fit to an extended Sellmeier model. The device pattern is created in positive-tone resist by electron-beam lithography. The pattern is then transferred to ${\rm Si_3N_4}$ by reactive ion etching using a ${\rm CF_4/CHF_3}$ chemistry. The device is chemically cleaned to remove deposited polymer and remnant resist, and then annealed at 1100~${\rm ^{\circ} C}$ in an ${\rm N_2}$ environment for 4 hours. An oxide lift-off process is performed so that the microrings have an air cladding on top while the input/output edge-coupler waveguides have ${\rm SiO_2}$ on top to form more symmetric modes for coupling to optical fibers. The facets of the chip are then polished for lensed-fiber coupling. After polishing, the chip is annealed again at 1100~${\rm ^{\circ} C}$ in an ${\rm N_2}$ environment for 4 hours.

\noindent \textbf{Translation Efficiency}
We assess the efficiency of spectral translation throughout the paper based on optical power. The translation efficiency $\eta$ is defined as $P_\text{SHG}/P_\text{pump}$ for second-harmonic generation (SHG, $\chi^{(2)}$), and as $P_\text{translated}/P_\text{signal}$ for other nonlinear processes, i.e., sum-/difference-frequency generation (SFG/DFG, $\chi^{(2)}$) and four-wave mixing (dFWM and FWM-BS, $\chi^{(3)}$). Here, $P$ represents the optical power on-chip and in the waveguide unless specified otherwise (see Supplementary Information for the itemized losses in the testing setup). The other figure of merit is quantum efficiency ($\eta_\text{Q}$), which is defined as $N_\text{SHG}/N_\text{pump}$ and $N_\text{translated}/N_\text{signal}$ for SHG and other processes, where $N$ represents photon flux on-chip and in the waveguide. In our work, the ratio of the translation efficiency and photon flux efficiency is 2.35, which equals the ratio of the photon energies at 669.8~nm and 1572.7~nm.

\bibliographystyle{osajnl}

\noindent \textbf{Acknowledgements} This work is supported by the DARPA DODOS and NIST-on-a-chip programs. X.L., G. M., Q.L., A.S., and A.R. acknowledge support under the Cooperative Research Agreement between the University of Maryland and NIST-CNST, Award no. 70NANB10H193.

\noindent \textbf{Additional Information} Correspondence and requests for materials should be addressed to X.L. and K.S.

\end{bibunit}

\newpage
\onecolumngrid \bigskip

\begin{bibunit}

\begin{center} {{\bf \large Supplementary Information}}\end{center}

\setcounter{figure}{0}
\makeatletter
\renewcommand{\thefigure}{S\@arabic\c@figure}

\setcounter{equation}{0}
\makeatletter
\renewcommand{\theequation}{S\@arabic\c@equation}

\section{Theoretical estimate}
In this section, we show that a $\chi^{(3)}$-based spectral translation process can be comparable to a $\chi^{(2)}$ process through proper nanophotonic mode engineering. We compare the translation efficiency of sum frequency generation (SFG) and degenerate four-wave mixing (dFWM) processes, the results of which were summarized in Fig.~1(d) in the main text. The nonlinear polarization ($\mathcal{P}$) of these two processes at the visible frequency ($\omega_v$) are given by\cite{Boyd2008}:
\begin{eqnarray}
\mathcal{P}^{(2)}(\omega_v)=2\epsilon_0\chi^{(2)}E(\omega_p)E(\omega_t),
\label{eq1}         \\
\mathcal{P}^{(3)}(\omega_v)=3\epsilon_0\chi^{(3)}E^2(\omega_p)E^*(\omega_t).
\label{eq2}
\end{eqnarray}
Here $\epsilon_0$ is the permittivity of free space, $E(\omega_i)$ is the electric field at frequency $\omega_i$, and $i=v,p,t$ represents visible, pump, and telecom wavelengths, respectively. First, we compare these two processes in the waveguide. The light intensity in the waveguide can be estimated as
$E(\omega_{p,t})=\sqrt{2I(\omega_{p,t})/(\epsilon c)}$ and $\mathcal{P}(\omega_v)=\sqrt{2I(\omega_v)/(\epsilon c)}$. We assume the material permittivity $\epsilon$ to be the same at all three wavelengths (accurate to first-order) and $c$ is the speed of light in vacuum. The intensity is given by $I=P/A$, where $P$ and $A$ represent optical power and effective mode area, respectively. We assume these two processes take place for the same set of visible/telecom modes and an ideal mode overlap. The ratio of the visible intensities for the two processes in a waveguide, as shown in the black line in Fig.~1(d), is thus given by:

\begin{eqnarray}
{Ratio={|\frac{\mathcal{P}^{(3)}(\omega_v)}{\mathcal{P}^{(2)}(\omega_v)}|}^2 = \frac{9}{2}|\frac{\chi^{(3)}}{\chi^{(2)}}|}^2 \frac{P_p}{\epsilon c A_p}.
\end{eqnarray}

In this equation, $\epsilon$ is the permittivity of the $\chi^{(3)}$ medium, and $P_p$ and $A_p$ represent the optical pump power in the waveguide and the effective mode area at the pump wavelength.

 The microring parameters we use for comparison are based on the typical geometry in this work (see Table~I in the main text for details), and the waveguide shares the same dimensions as the microring cross-section. The wavelengths of interest are 1560~nm and 780~nm for the input signal and translated signal, respectively. Nonlinear refractive indices used in the comparison are ${\rm \chi^{(3)}=3.39\times10^{-21}~m^2/V^2}$ (${n_{2}=2.4\times10^{-15}{\rm cm^2/W}}$) and ${\rm \chi^{(2)}=60~pm/V}$, which are typical values for ${\rm Si_3N_4}$ (Ref.~\onlinecite{Ikeda2008}) and ${\rm LiNbO_3}$ (Ref.~\onlinecite{Boyd2008}), respectively. We note that many ${\rm LiNbO_3}$ photonic works quote smaller $\chi^{(2)}$ values due to technical issues (thin film quality, quasi-phase-matching, etc.), but we use the largest bulk nonlinearities for ${\rm LiNbO_3}$ to give a conservative comparison. The effective area for the pump mode is 0.18~${\rm \mu m^2}$, simulated by the finite-element method. While the modal confinement provided by a waveguide provides an enhancement over bulk material, the pump power required for a  $\chi^{(3)}$ process to be comparable to $\chi^{(2)}$ process is unreasonably large (black line in Fig.~1(d)), and is off the $x$-axis range displayed.

A high-$Q$ microring provides strong confinement of light in both space and time, resulting in greatly enhanced optical intensities. The enhancement of the optical intensity in the microring can be estimated by its finesse (${\mathcal{F}=Q \frac{FSR}{\omega/2\pi}}$). The efficiency ratio $\eta[\chi^{(3)}]/\eta[\chi^{(2)}]$ is thus multiplied by ${\rm \mathcal{F}}$, as shown by the blue line in Fig.~1. The assumed finesse ($\mathcal{F}$) of the microring is 5000, which corresponds to an optical quality factor of ${Q=10^6}$ for a free spectral range ($FSR$) of 1~THz, consistent with measurements for our device. With the cavity enhancement, the efficiency ratio $\eta[\chi^{(3)}]/\eta[\chi^{(2)}]$ is much larger than it is in a waveguide, but is still only -44~dB at 1~mW. However, for the estimates considered so far, we have assumed perfect mode overlap among the interacting modes, whereas in practice the the mode overlap of the $\chi^{(2)}$ process has usually been less than ideal, while in our $\chi^{(3)}$ device the overlap is close to unity. Taking this factor into account, we have a further enhancement of the efficiency ratio, as shown by the red line in Fig.~1(d). The mode overlap of the $\chi^{(2)}$ process is estimated to be 0.3, which is typical because different and higher-order mode families are used for $\chi^{(2)}$ dispersion matching\cite{Ilchenko2004}. An extra factor of 0.043 takes into account the difficulties to identify/employ the targeted modes, which has especially been a problem for mm-scale devices in which many mode families are supported~\cite{Furst2010}. We note that this extra factor could ideally be removed if optical modes of the same family can be unambiguously identified and frequency-matched and phase-matched in a nanophotonic $\chi^{(2)}$ process. However, this has not been achieved in the context of wide-band nanophotonic spectral translation, to the best of our knowledge.

Although we take SFG and dFWM as examples here, the comparison is generally applicable to $\chi^{(2)}$ and $\chi^{(3)}$ processes. For SHG and non-degenerate FWM, Eq.~\ref{eq1} and Eq.~\ref{eq2} are corrected by a factor of 1/2 and 2, respectively, due to the degenerate/non-degenerate nature of the processes. The estimate of efficiency ratio is therefore within a factor of 4 depending on the exact processes being compared, which does not affect the general trends shown in Fig.~1.

\section{Experimental Setup}
\begin{center}
\begin{figure*}[htbp]
\begin{center}
\includegraphics[width=1.0\linewidth]{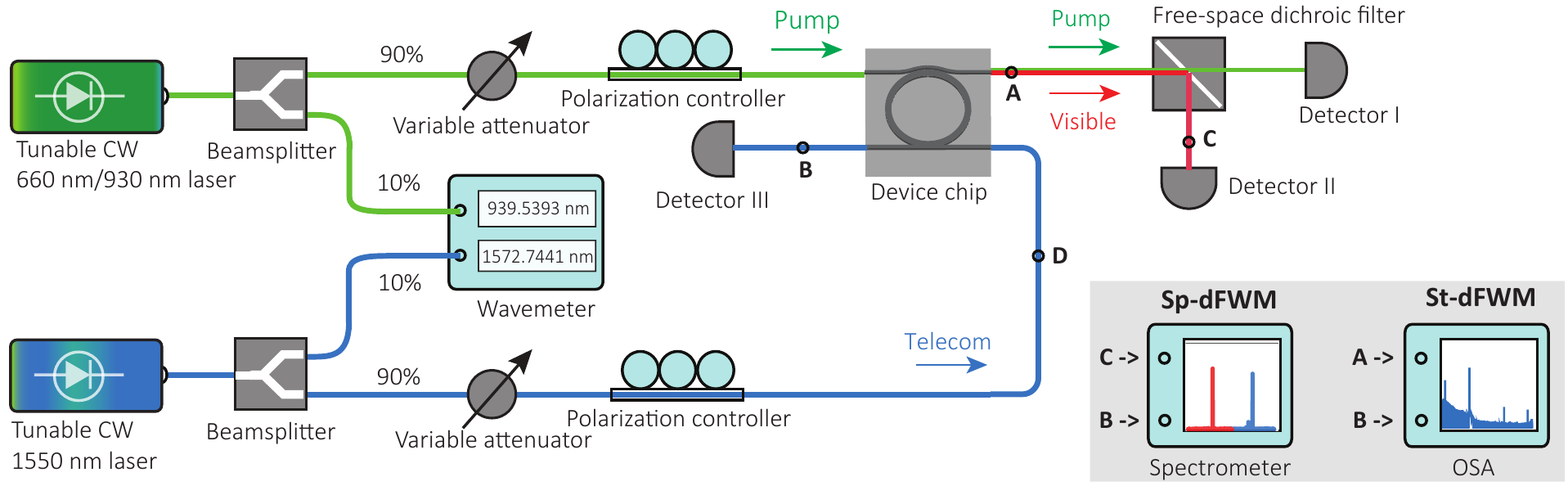}
  \caption{\textbf{Experimental Setup}. Figure of the experimental setup used to record cavity transmission, photon pair spectra by Sp-dFWM, and spectral translation by St-dFWM. CW: continuous-wave. In the Sp-dFWM case, only the 930~nm (pump) laser is used. In the St-dFWM case, 930~nm (pump) and 1550~nm (telecom) lasers are used. Points A-D are used to guide the different configurations in the experiments. Sp-dFWM: spontaneous degenerate four-wave mixing. St-dFWM: stimulated degenerate four-wave mixing.}\label{Fig_S1}
\label{fig:FigS1}
\end{center}
\end{figure*}
\end{center}

The device is tested with the experimental setup shown in Figure~\ref{fig:FigS1}, which illustrates the measurement configurations for cavity transmission, photon pair spectrum of spontaneous degenerate four-wave mixing (Sp-dFWM), and spectral translation by stimulated degenerate four-wave mixing (St-dFWM). For the cavity transmission measurement, three tunable continuous-wave lasers are used in turn to measure the transmission spectra of the signal, pump, and idler bands. The wavelengths of the cavity modes are calibrated by a wavemeter with an accuracy of 0.1~pm. The pump laser is attenuated to sub-milliwatt levels to avoid thermal bistability with the polarization adjusted to transverse-electric (TE). Pump, visible, and telecom signals are coupled on and off the chip by lensed optical fibers with a focused spot size diameter of ${\rm \approx2.5~\mu}$m. To achieve mode profiles that best match to these fibers, the input/output waveguides are tapered to expand their optical modes, with an oxide lift-off process to make the modes symmetric in the vertical direction. The measured fiber-chip insertion losses are (3.33~$\pm$~0.24)~dB, (2.92~$\pm$~0.15)~dB, and (2.97~$\pm$~0.09)~dB per facet at 669.8~nm, 939.5~nm, and 1572.7~nm, respectively. The uncertainty is calculated as a one standard deviation value from the transmission background (Fig.~3(a) in the main text). These insertion losses are used to calibrate the on-chip pump/telecom power from the measured power in the fiber that is incident on the chip. In measurements of the device's spectrum, no filters are required before the device, since both the laser's amplified-spontaneous-emission noise and the Raman noise generated in the optical fibers have a frequency spectrum that is far from the visible band. When measuring the spectrum of Sp-dFWM and calibrating power-dependent St-dFWM measurements, a free-space dichroic setup is used. The dichroic setup contains fiber-to-free-space coupling, a 950~nm dichroic mirror, two cascaded bandpass filters at 670~nm with 10~nm full-width at half-maximum (FWHM) (providing $>$80~dB isolation), and free-space-to-fiber coupling for detection. The overall loss for the 670~nm path of the dichroic filter is (3.44$~\pm~$0.06)~dB. Since this wavelength is not reachable by our red laser, the loss is calibrated with the optical spectrum analyzer (OSA) by using the spectrally translated signal before/after the dichroic filter. The uncertainty represents one-standard deviation from six measurements, which is likely due to uncertainty from fiber connection and/or OSA detection.

\end{bibunit}

\end{document}